\documentstyle[epsfig]{aipproc}

\def\sb{\ifmmode{\;{\rm mag}\;{\rm arcsec}^{-2}}\else{~mag~arcsec$^{-2}$}\fi}
\def\csb{\ifmmode{\mu_0}\else{$\mu_0$}\fi}

\begin{document}
\title{Gas Content and Star Formation \\
	Thresholds in the Evolution of \\
	Spiral Galaxies}
\author{Stacy McGaugh$^*$ and Erwin de Blok$^{\dagger}$}
\address{$^*$Department of Terrestrial Magnetism,
Carnegie Institution of Washington \\
$^{\dagger}$Kapteyn Astronomical Institute, University of Groningen}

\lefthead{McGaugh \& de Blok}
\righthead{Gas and Star Formation in Spiral Galaxies}
\maketitle

\begin{abstract}

The gas mass fraction of spiral galaxies is strongly
correlated with the central surface brightness of their disks.
There exist many dim galaxies with long gas consumption time scales
and $f_g > 0.5$.  This resolves the gas consumption paradox.

The surface density of gas follows the optical surface brightness,
but does not vary by as large a factor.  This is the signature of
a critical density threshold for star formation.  Such a mechanism
seems to be responsible for the slow evolution of dim galaxies.
 
\end{abstract}

\section*{Gas Content}

The fraction of baryonic mass which has been converted from gas into stars
is a fundamental measure of the degree of evolution of a galaxy.  The gas
content of spiral galaxies is a strongly correlated with the optical
surface brightness of their disks (Fig.~1).  This must be an evolutionary
effect indicative of the rate of galaxy evolution in disks of differing
surface mass densities \cite{dBM1,hos}.

A complete analysis describing the details of the derivation of
the gas mass fractions is given elsewhere \cite{MdB}.
Gas content increases strongly with decreasing surface brightness.
An important consequence of this is the end of the gas consumption
paradox.  Gas rich galaxies do exist, and are quite common \cite{me}.

\section*{Disk Evolution}

The present epoch gas fraction is a direct chronometer of the star formation
history.  There is degeneracy in how a given $f_g$ may be reached,
but the evolutionary rate and/or age of spiral galaxies must
vary systematically with surface brightness to give the observed correlation.
Burst and fade scenarios \cite{Pad} can be completely ruled out as these should
result in low not high $f_g$ for dim galaxies.  Indeed, it is difficult
to have exponentially declining star formation histories in galaxies with
$f_g > 0.5$ unless they are quite young.  Such objects can be old if
the star formation rate increases rather than decreases with time,
but only at the expense of
making the mean age of the stars very young (a few Gyr).  A roughly constant
star formation rate is more plausible, but requires an intermediate age
(8 or 10 Gyr rather than 12 or 14 Gyr).

Galaxies with $f_g > 0.5$ have most of their star forming
potential in the {\it future}.  These gas rich galaxies are inevitably
morphologically late types (Sd \& Sm).  They could not possibly have
experienced an evolution of rapid gas consumption followed by
rapid fading.  Yet this is precisely the evolution inferred for late types
from high redshift data \cite{Mad}.  These results are both sound
and utterly contradictory.  The nature of the faint blue galaxies
therefore remains a mystery.

\section*{Star Formation}

Why have dim galaxies converted so little of their gas into stars?

A first approximation of the dependence of the star formation rate on
gas density is the Schmidt Law:
\begin{equation}
\dot f_g \propto \Sigma_g^N.
\end{equation}
Kennicutt \cite{K89} finds $N = 1.3 \pm 0.3$.
By this criterion, low surface density disks should evolve slowly,
but this alone is not adequate to explain the observations.
Gas surface density does follow optical surface brightness, but does
not vary by as large a factor \cite{dB2}.  Roughly speaking,
$\Sigma_g$ drops by a factor of 2 for a factor of 5 drop in \csb.
Making the usual assumption that the amount of light
traces the global star formation rate averaged over a Hubble time,
\begin{equation}
\frac{\Sigma_{opt}(HSB)/T_H}{\Sigma_{opt}(LSB)/T_H} \approx
\frac{\dot f_g({\rm HSB})}{\dot f_g({\rm LSB})} =
\left[\frac{\Sigma_g({\rm HSB)}}{\Sigma_g({\rm LSB})}\right]^N \rightarrow
5 = 2^N
\end{equation}
requires a value of $N \approx 2.3$, much larger than observed.
Though this is a crude calculation, it is a conservative one:  other
plausible assumptions require even larger $N$.

\begin{figure}[fgBI.ps] 
\centerline{\epsfig{file=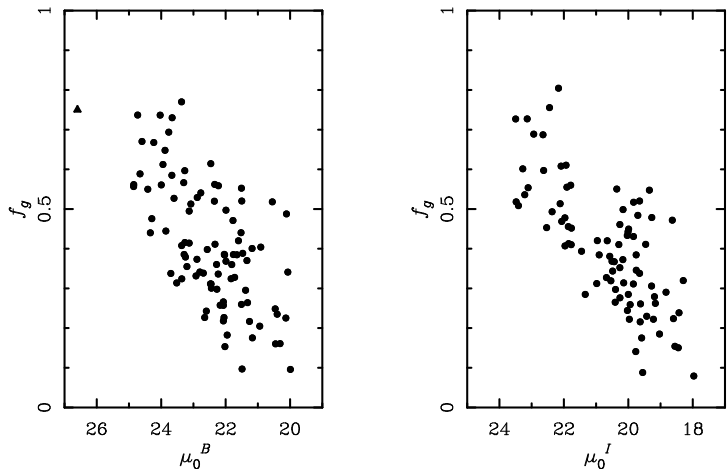,height=2.5in}}
\vspace{10pt}
\caption{The gas mass fraction as a function of disk central surface
brightness.  The two panels show results derived independently from a)
$B$-band and b) $I$-band data.  These give consistent results.  The
slopes differ slightly because higher surface brightness galaxies tend to
be redder.  The data show a strong correlation [${\cal R} = 0.63$ in (a)]
with gas fraction decreasing as surface brightness increases.
This goes in the sense expected if low densities inhibit star formation.}
\label{fig1}
\end{figure}

\begin{figure}[f563v1.ps] 
\centerline{\epsfig{file=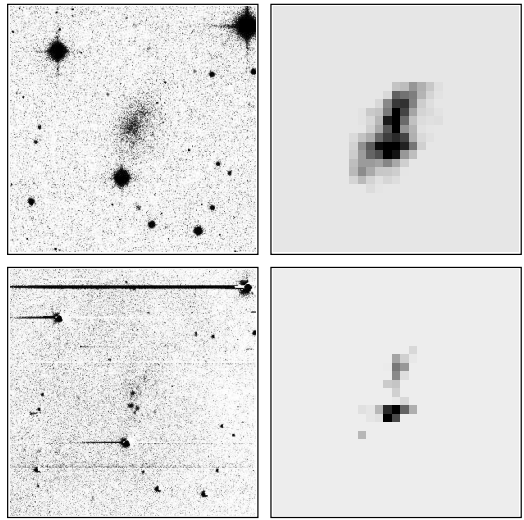,height=2.7in}}
\vspace{10pt}
\caption{The low surface brightness galaxy F563--V1.  a) Broad band optical
($V$-band) image.  b) HI gas distribution.  c) Continuum subtracted H$\alpha$
emission.  d) Critical density map.  The critical density map is constructed
by dividing the gas distribution in (b) by $\Sigma_c$ as given in the text.
Only regions near to or exceeding this threshold are shown.  There is a
reasonable correspondence between these areas which should be forming stars
and those which actually are in (c).}
\label{fig2}
\end{figure}

The next order approximation for star formation is based on
local gravitational stability in the disk \cite{K89}.
This leads to a critical density threshold below which star formation
is suppressed:
\begin{equation}
\Sigma_c = \frac{\sigma}{3.4 G}\frac{V}{R}\left(1+\frac{R}{V}
\frac{dV}{dR}\right)^{1/2}.
\end{equation}
This provides a good explanation for the evolutionary
sluggishness of low surface brightness disks:  they are at or below the
critical threshold.

We have tested this in a radially averaged way \cite{vdH}
and it works well:  lower surface brightness disks are generally further
into the critical regime.  We can also test this formulation on
a point by point basis (Fig.~2).
In many cases it works well, there being
a reasonable coincidence between regions where $\Sigma_g > \Sigma_c$
and the location of HII regions.

Nevertheless, the agreement between prediction and observation is by
no means perfect in all cases.  The $\Sigma_c$ criterion seems especially
prone to failure in the solid-body portion of the slowly rising rotation
curves of low surface brightness galaxies.  This is perhaps not surprising;
there is very little shear in these regions.
Stability is maintained for lower $Q$ values, and asymmetries can persist in
the gas for many dynamical times.  Nevertheless, star formation is sometimes
observed in regions which should be stable.  It seems that further physics
is at work --- both the Schmidt law and Kennicutt's critical density criteria
are operative and directly relevant to the global evolution of galaxies,
but they are not the end of the story.

Though further physics is required to understand all the details of star
formation, one basic result is clear.  Surface mass density is a
critical parameter in determining the evolutionary rate of disks.
Dim galaxies have consumed little of their gas because of their low surface
densities.

\section*{Galaxy Formation}

If surface density determines the evolution of disk galaxies,
what determines the surface density?
The origin of disk surface density is probably related to the
amplitude of the density fluctuation from which a galaxy was born.
Lesser density perturbations will
expand longer before turn around, and collapse later.  This results
in a lower final mass concentration and a younger age, as observed.
Density begets density.


\begin{references}

\bibitem{dBM1} de Blok, W.J.G., \& McGaugh, S.S. 1996, ApJ, 469, L89
	 
\bibitem{hos} Mihos, J.C., McGaugh, S.S., \& de Blok, W.J.G. 1997,
	these proceedings

\bibitem{MdB} McGaugh, S.S., \& de Blok, W.J.G. 1997, ApJ, in press

\bibitem{me} McGaugh, S.S. 1996, MNRAS, 280, 337

\bibitem{Pad} Padoan, P., Jimenez, R., \& Antonuccio-Delogu, V.
	1997, these proceedings \& astro-ph/9609091

\bibitem{Mad} Madau, P. 1997, these proceedings

\bibitem{K89} Kennicutt, R.C. 1989, ApJ, 344, 685

\bibitem{dB2} de Blok, W.J.G., McGaugh, S.S., \& van der Hulst, J.M.
		   1997, MNRAS, 283, 18
		    
\bibitem{vdH} van der Hulst, J.M., Skillman, E.D., Smith, T.R., Bothun, G.D.,
McGaugh, S.S., \& de Blok, W.J.G. 1993, AJ, 106, 548

\end{references}
\end{document}